\documentstyle[12pt,epsf]{article}
\begin{document}
%%%%%%%%%%%%%%%%%%%%%%%%%%%%%%%%%%%%%%%%%%%%%%%%%%%%%%%%%%%%%%%%%%%%%
\newsavebox{\tadpole}
\savebox{\tadpole}{
\begin{picture}(0,0)
\unitlength1mm
\put(5,0){\circle{6}}
\put(5,0){\circle{5}}
\put(0,-3){\line(1,0){10}}
\end{picture} }
%%%%%%%%%%%%%%%%%%%%%%%%%%%%%%%%%%%%%%%%%%%%%%%%%%%%%%%%%%%%%%%%%%%%%
\newsavebox{\basketball}
\savebox{\basketball}{
\begin{picture}(0,0)
\unitlength1mm
\put(5,0){\circle{6}}
\put(5,0){\circle{5}}
\put(5,0){\oval(10,6)}
\put(5,0){\oval(11,7)}
\end{picture} }
%%%%%%%%%%%%%%%%%%%%%%%%%%%%%%%%%%%%%%%%%%%%%%%%%%%%%%%%%%%%%%%%%%%%%
\newsavebox{\sunrise}
\savebox{\sunrise}{
\begin{picture}(0,0)
\unitlength1mm
\put(5,0){\circle{6}}
\put(5,0){\circle{5}}
\put(0,-.1){\line(1,0){10}}
\put(2.5,.2){\line(1,0){5}}
\end{picture} }
%%%%%%%%%%%%%%%%%%%%%%%%%%%%%%%%%%%%%%%%%%%%%%%%%%%%%%%%%%%%%%%%%%%%%
\newsavebox{\vertex}
\savebox{\vertex}{
\begin{picture}(0,0)
\unitlength1mm
\put(5,0){\circle{6}}
\put(5,0){\circle{5}}
\put(0,-3){\line(1,0){10}}
\put(5,-2.75){\circle*{2}}
\end{picture} }
%%%%%%%%%%%%%%%%%%%%%%%%%%%%%%%%%%%%%%%%%%%%%%%%%%%%%%%%%%%%%%%%%%%%%
\newsavebox{\doublecircle}
\savebox{\doublecircle}{
\begin{picture}(0,0)
\unitlength1mm
\put(5,0){\circle{6}}
\put(5,0){\circle{5}}
\put(10.5,0){\circle{6}}
\put(10.5,0){\circle{5}}
\end{picture} }
%%%%%%%%%%%%%%%%%%%%%%%%%%%%%%%%%%%%%%%%%%%%%%%%%%%%%%%%%%%%%%%%%%%%%

% \vspace*{3cm} % submission

\begin{center}
{\large \bf Thermodynamics of the \hspace{-2.7mm}
$\mbox{\boldmath{$\phi^4$}}$ theory
in tadpole approximation}\\[6mm]
{\sc A. Peshier$^a$, B. K\"ampfer$^a$, O.P. Pavlenko$^{a,b}$,
G. Soff$^c$}\\[6mm]
$^a$Research Center Rossendorf, PF 510119, 01314 Dresden, Germany\\
$^b$Institute for Theoretical Physics, 252143 Kiev - 143, Ukraine\\
$^c$Institut f\"ur Theoretische Physik, TU Dresden, 01065 Dresden, Germany
\end{center}

\vspace*{6mm}

\centerline{{\bf Abstract}}
Relying on the Luttinger-Ward theorem we derive a thermodynamically
selfconsistent and scale independent approximation of the thermodynamic
potential for the scalar $\phi^4$ theory in the tadpole approximation.
The resulting thermodynamic potential as a function of the temperature
is similar to the one of the recently proposed screened perturbation theory.

%\newpage       % submission
\vspace*{2cm} % preprint

The usual perturbative evaluation of the QCD thermodynamic potential
$\Omega(T)$ up to fifth order in the coupling constant $g$
\cite{5_QCD} encounters the problem that the contributions
$\propto g^2$ to $g^5$ alternate in sign and increase in magnitude and
do not seem to approach a limiting value in
the strong coupling regime. Therefore the hope has been formulated
that a suitably reorganized perturbation theory can improve the calculation
of $\Omega$. This problem has been attacked in refs.~\cite{Karsch,Rebhan}.
For instance, in ref.~\cite{Karsch}
a screened loop expansion is employed which accounts for the screening
in thermal propagators. Such an improvement of the convergence
properties in analytical calculations
of $\Omega$ is urgently necessary since recent lattice QCD calculations
\cite{QCD_lattice} have delivered accurate numerical
results which now should be understood on a
quantitative level in the physically
relevant region, say up to temperatures of $5 \, T_c$, where significant
deviations from the asymptotic limit of $\Omega \propto T^4$ are still
clearly seen. Near the confinement temperature $T_c$ these
deviations are, of course, very large. One can describe
such a behavior on the basis of effective theories
with parameters adjusted to
the lattice results (see \cite{Peshier,Levai} for recent attempts and
references quoted therein), but most {\sl ans\"atze} are hampered by their
{\sl ad hoc} character or vague contact to QCD as the fundamental
microscopic gauge field theory.

Since QCD has a quite complicated analytical structure, one should start to
develop a systematic approach for simpler model field theories.
Here we consider the scalar massless $\phi^4$ theory with the Lagrangian
${\cal L} = \frac12 (\partial_\mu \phi) (\partial^\mu \phi)
- \frac{g^2}{4!} \phi^4$.
For such a model the perturbative expansion of $\Omega$
is also performed up to
fifth order \cite{5_phi}, and it shows similar bad convergence properties
as QCD at not too small coupling strength $g$.
We are going to derive from the
Luttinger-Ward theorem an approximation of $\Omega(T)$.
The approach of Luttinger and Ward is known to be rather general and
allows for selfconsistent approximations respecting the symmetries and
conservation laws of the underlying full theory.

The Luttinger-Ward theorem can be formulated for the scalar theory as
\begin{eqnarray}
\Omega & = & \frac 12 T V \sum \hspace*{-5.0mm} \int \;
\{\ln (- \Delta^{-1}) + \Delta \Pi \} + \Omega',\\
\Omega' & = & - \sum_n \frac{1}{4 n} T V \sum \hspace*{-5.0mm} \int \;
\Delta \Pi^s_{(n)},
\end{eqnarray}
where $\Delta$ denotes the exact propagator and $\Pi$ the exact selfenergy;
$\Pi^s_{(n)}$ stands for the contribution of the skeleton diagrams of
$n$th order to the selfenergy (these are two-particle irreducible diagrams);
$V$ is the volume of the system,
and $\small{\vspace*{-3mm}\Sigma}\!\!\!\!\!\int$ is explained below.
$\Omega$ has the property of stationarity \cite{Li_Yan}, i.e.,
\begin{equation}
\frac{\delta \Omega}{\delta \Delta} = 0,
\label{selfcon}
\end{equation}
which ensures the selfconsistent
relation between $\Omega$, $\Delta$ and $\Pi$.
Originally the relations (1, 2) have been derived for
Fermi systems \cite{LW}, later it has been extended to Bose systems
\cite{Fulde}. For details of the derivation of the  formulation (1, 2)
we refer the interested reader to \cite{Norton_Corn,Peshier_PhD}.

Here we use eqs.~(1, 2) to derive a consistent approximation of $\Omega$.
Our starting point is the restriction to the tadpole contribution to the
selfenergy. This implies that we take into account only the first term
in the functional $\Omega'$, i.e.,
\begin{equation}
\Omega'_{(1)} = - \frac 14 T V \sum \hspace*{-5.0mm} \int \;
\Delta_{(1)} \, \Pi^s_{(1)} = - 3
\put(0,3){\usebox{\doublecircle}}
\end{equation}
($\Delta_{(1)}$ and
double lines in the diagrammatic representation denote the full
propagator in the considered approximation).
The corresponding selfenergy is due to
eq.~(\ref{selfcon})
\begin{equation}
\Pi^s_{(1)} = 12 \left(\frac{-g^2}{4!} \right)
T \sum \hspace*{-5.0mm} \int \;
\Delta_{(1)} = 12
\put(0,3){\usebox{\tadpole}} \hspace*{10mm} \hspace*{3mm} .
\end{equation}
In the tadpole approximation the selfenergy is momentum independent
and real. Therefore, in this approximation the corresponding resummed
propagator looks like the free propagator modified by a term which
resembles a mass term, i.e.,
$\Delta^{-1} \to \Delta^{-1}_{(1)} = \Delta^{-1}_{(0)} - \Pi_{(1)}^s$,
and $\Pi_{(1)}^s$ is equivalent to the 1-loop selfenergy.

Since after performing the Matsubara sum the integrals over loop momenta
are ultraviolet divergent
one has to regularize. We employ the $\overline{\mbox{MS}}$ scheme in
$4 - 2 \epsilon$ dimensions, where
\begin{equation}
\sum \hspace*{-5.0mm} \int \; =
\left( \frac{e^\gamma \bar \mu^2}{ 4 \pi} \right)^\epsilon
\sum_{p_0} \int \frac{d^{3-2\epsilon} p}{(2\pi)^{3-2\epsilon}};
\end{equation}
$\bar \mu$ is the renormalization scale, and $\gamma \approx 0.577$ is
Euler's number. An explicit calculation of eq.~(5) yields
\begin{eqnarray}
\Pi^s_{(1)} & = & 12 \left( -\frac{g^2}{4!} \right)
\left\{
\frac{\Pi^s_{(1)}}{(4\pi)^2}
\left[
\frac1\epsilon + \ln \frac{\bar \mu^2}{T^2} -
\ln \frac{\Pi^s_{(1)}}{T^2} +1
\right]
\right. \nonumber \\
&& -
\left.
\frac{1}{2\pi^2} \int_0^\infty dp \,
\frac{p^2}{\sqrt{p^2 + \Pi^s_{(1)}}}\,
\frac{1}{ \exp \left\{ \frac{ \sqrt{p^2 + \Pi^s_{(1)}} }{T} \right\} - 1}
+ {\cal O} (\epsilon)
\right\}
\label{x}
\end{eqnarray}
which is an implicit equation for $\Pi^s_{(1)}$.
The first term $\propto \epsilon^{-1}$ corresponds to
the divergent vacuum part in the usual vacuum selfenergy of the massive
scalar theory.
Since now $\Pi^s_{(1)}$ is temperature dependent, also this term is
temperature dependent.
Therefore, it cannot be cancelled by vacuum counter terms.
It can be shown, however, that this term as well as the scale dependent
term $\propto \ln (\bar \mu^2 / T^2)$ are cancelled by higher loop
contributions.
This can be seen explicitly when considering all contributions
$\propto g^4 T^2 (\frac 1\epsilon + \ln \frac{\bar \mu^2}{T^2})$
to the exact selfenergy.
In our case the relevant contributions come from the above tadpole,
$12
\put(0,3){\usebox{\tadpole}} \hspace*{12mm}
\to 12 \left( - \frac{g^2}{4!} \right)
\frac{\Pi^s_{(1)}}{(4\pi)^2} (\frac 1\epsilon + \ln \frac{\bar \mu^2}{T^2})$,
the rising sun diagram,
$96
\put(0,3){\usebox{\sunrise}} \hspace*{12mm}
\to -96 \left( - \frac{g^2}{4!} \right)^2 \frac{T^2}{(4\pi)^2}
\frac14 (\frac 1\epsilon + \ln \frac{\bar \mu^2}{T^2}) $,
and the vertex correction in the tadpole,
$12
\put(0,3){\usebox{\vertex}} \hspace*{12mm}
\to -12 \left( - \frac{g^2}{4!} \right)
\frac{3 g^2}{(4 \pi)^2} \frac{T^2}{12}
\frac12 (\frac 1\epsilon + \ln \frac{\bar \mu^2}{T^2})$,
which indeed cancel since in leading order
$\Pi^s_{(1)} \to g^2 T^2 /4!$.
The regularized part of eq.~(\ref{x}) thus allows the identification
$\Pi^s_{(1)} \to m^2$ and reads at $\epsilon \to 0$ like a gap equation
\begin{equation}
m^2 =  \frac{g^2}{2} \left\{
\frac{1}{2\pi^2} \int_0^\infty dp \,
\frac{p^2}{\sqrt{p^2 + m^2}}
\frac{1}{\exp \left\{ \frac{\sqrt{p^2 + m^2 }}{T} \right\} - 1}
+
\frac{m^2}{(4\pi)^2} \left[
\ln \frac{m^2 }{T^2} - 1 \right]
\right\} ,
\label{xx}
\end{equation}
which determines $m(T)$.
Therefore, in our approach the gap equation is inherent, while
in ref.~\cite{Karsch} it is taken as an external information.
With the definition $\Delta_{m^2}^{-1} = \Delta_0^{-1} - m^2$
our approximation of the thermodynamic potential is determined by
\begin{eqnarray}
\Omega_{(1)} & = &
\frac12 T V \sum \hspace*{-5.0mm} \int \;
\left\{ \ln (-\Delta^{-1}_{m^2}) + \frac12 m^2 \Delta_{m^2}
\right\} \label{y} \\
& = & \frac{T V}{2 \pi^2} \int_0^\infty dp \, p^2
\ln \left(
1 - \exp \left\{ - \frac{\sqrt{p^2 + m^2}}{T} \right\} \right) \nonumber\\
&& -
\frac{m^2 V}{8 \pi^2}
\int_0^\infty dp \, p^2
\frac{1}{\sqrt{p^2 + m^2}}
\left[ \exp \left\{ \frac{\sqrt{p^2 + m^2}}{T}  \right\}
- 1 \right]^{-1}
- \frac{m^4}{128 \, \pi^2} ,
\label{eos}
\end{eqnarray}
where we used $\partial_{m^2} \ln ( - \Delta^{-1}_{m^2}) = - \Delta_{m^2}$
for calculating
\begin{eqnarray}
T \sum \hspace*{-5.0mm} \int \; \ln (-\Delta^{-1}_{m^2} )
& = &
- \frac 12 \frac{m^4}{(2\pi)^2} \left[
\frac 1 \epsilon + \ln \frac{\bar \mu^2}{T^2} - \ln \frac{m^2}{T^2}
+\frac 32 \right]  \nonumber \\
&& +
\frac{T}{\pi^2} \int_0^\infty dp \, p^2
\ln \left( 1 - \exp \left\{ -\frac{\sqrt{p^2 + m^2}}{T} \right\} \right)
+ {\cal O}(\epsilon). \nonumber
\end{eqnarray}
It is remarkable that in the expression (\ref{y}) the divergencies and
scale dependent terms exactly cancel and, in contrast to eq.~(\ref{x}),
no assumptions on cancelling higher-order contributions or so are needed.
In this sense, $\Omega_{(1)}$ is to be considered as a consistent
approximation.

It should be noted that the thermodynamic potential eq.~(\ref{eos}) looks
like the one for an ideal gas with temperature dependent thermal mass of
the particles (first term) plus two correction terms (second line).
In fig.~1 we show the numerical results of the solution of
eqs.~(\ref{xx},~\ref{eos}) for the pressure $p = - \Omega_{(1)}/V$ as a
function of the coupling strength.
Interestingly, our pressure is very similar to the one derived in the
screened loop expansion \cite{Karsch} up to $g = 4$;
at larger values of $g$ our pressure is somewhat larger.
The massive ideal gas
term alone turns out as a quite acceptable approximation
of the full expression eq.~(\ref{eos}) on the 10\% level.
The entropy density $s = \partial p / \partial T$
is exactly given by the formula for
the ideal gas of particles with mass $m(T)$, while the energy
density $e = sT - p$ consists of the massive ideal gas contribution
plus two correction terms (second line in eq.~(\ref{eos}) but with
opposite signs).
The form of the thermodynamical quantities is intimately related to the
fact that the excitations with mass $m(T)$ represent stable quasiparticles.
From the above one can conclude that the phenomenological approaches,
which use ideal gas approximations with thermal masses
\cite{Peshier,Levai,Karsch_89} for describing lattice data, get some support.

Going beyond the tadpole approximation means inclusion of the rising sun
diagram\\
$\put(-1,3){\usebox{\sunrise}}$ \hspace*{11mm}
in the selfenergy or fixing the functional
$\Omega'$ by
$\Omega'_{(2)} = \Omega'_{(1)}
% -3 \put(-3,3){\usebox{\doublecircle}} \hspace*{15mm}
- 12
\put(2,3){\usebox{\basketball}} \hspace*{14mm} . $
Here one gets a momentum dependent selfenergy which also possesses an
imaginary part. Then the sum integrals
%like $\sum$ \hspace*{-7mm} {\Large $\int$}
are not longer analytically calculable as before and deserve new techniques.

In QCD even more involved structures appear
since already at the 1-loop level the
selfenergies become momentum dependent and contain imaginary parts;
different mass scales also appear. Therefore, a direct application
of our scheme to QCD needs a more sophisticated approach.

In summary, basing on the Luttinger-Ward theorem for a selfconsistent
relation between thermodynamic potential, propagator, and selfenergy
we derive an expression for the thermodynamic potential
in the tadpole approximation for the $\phi^4$ theory which turns out
to be numerically similar to the one in the screened loop expansion.
The previous trails to approximate the thermal field theory in the
strong coupling regime by an effective ideal gas expression
for the entropy density of particles with
finite thermal mass are supported.

{\bf Acknowledgments:}
Useful discussions with
F. Karsch, R. Pisarski, H. Satz, and M. Thoma
are gratefully acknowledged.
This work is supported in part by BMBF grant 06DR829/1 and GSI.
%%%%%%%%%%%%%%%%%%%%%%%%%%%%%%%%%%%%%%%%%%%%%%%%%%%%%%%%%%%%%%%%%%%%%

%%%%%%%%%%%%%%%%%%%%%%%%%%%%%%%%%%%%%%%%%%%%%%%%%%%%%%%%%%%%%%%%%%%%%

\newpage

\begin{figure}
 \vskip 3cm
 \centerline{\epsfxsize=.7 \hsize
 \epsffile{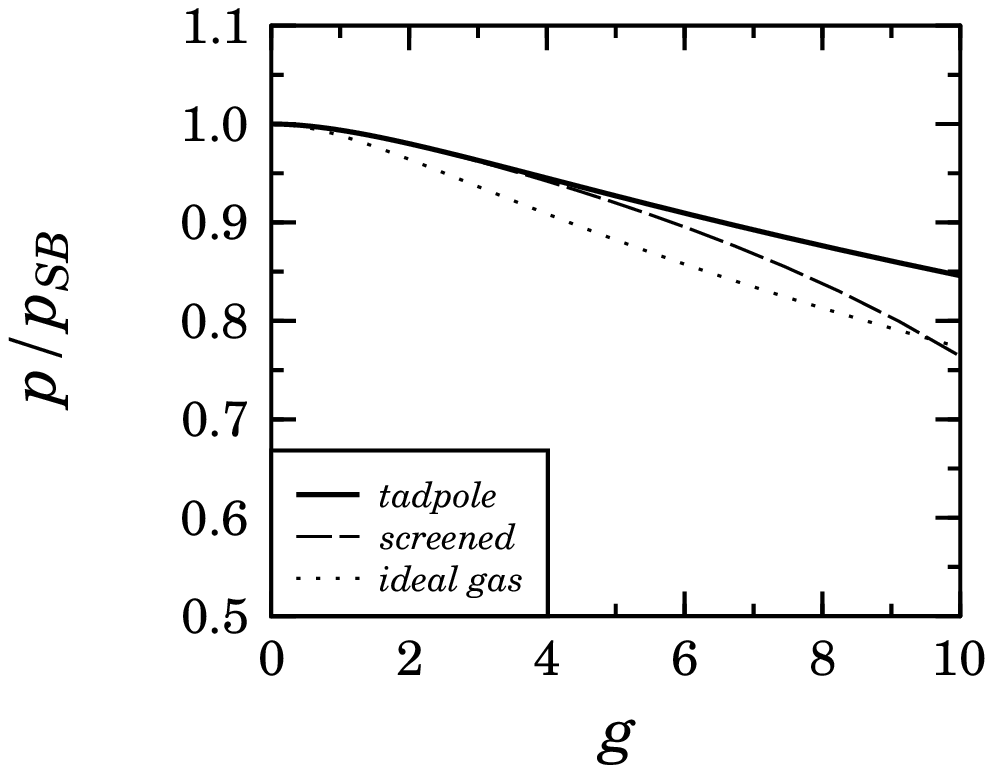}}
 \vskip 1cm
\caption{
The pressure (scaled by the Stefan-Boltzmann pressure
of an massless ideal gas) in our approach
eq.~(\protect\ref{eos}) (full line) as a function of the coupling strength.
The dotted line is the ideal massive gas contribution
(first term in eq.~(\protect\ref{eos})),
while the dashed line depicts the result
of the screened loop expansion \protect\cite{Karsch}.
}
\label{exp28}
\end{figure}

\end{document}